\documentstyle[prd,aps]{revtex}
\begin{document}
\draft
\input epsf
\def\la{\mathrel{\mathpalette\fun <}}
\def\ga{\mathrel{\mathpalette\fun >}}
\def\fun#1#2{\lower3.6pt\vbox{\baselineskip0pt\lineskip.9pt
        \ialign{$\mathsurround=0pt#1\hfill##\hfil$\crcr#2\crcr\sim\crcr}}}

\twocolumn[\hsize\textwidth\columnwidth\hsize\csname
@twocolumnfalse\endcsname

\title{An Inflaton Candidate in Gauge Mediated Supersymmetry Breaking}
\author{Micheal Dine$^{(1)}$ and Antonio
Riotto$^{(2)}$}
 \address{$^{(1)}${\it Santa Cruz Institute for Particle Physics, University of California, Santa Cruz, CA~~95064}}
\address{$^{(2)}${\it NASA/Fermilab Astrophysics Center, \\ Fermilab
National Accelerator Laboratory, Batavia, Illinois~~60510-0500}}
\date{May 15, 1997}
\maketitle
\begin{abstract}
Inflation, as currently understood, requires the presence of fields with
very flat potentials.  Supersymmetric models in which
supersymmetry breaking is communicated by supergravity naturally yield such fields,
but the scales are typically not suitable for obtaining both
sufficient inflation and a suitable fluctuation spectrum.  In the context
of recent ideas about gauge mediation, there are new candidates for the inflaton.
We present a simple model for slow-rollover inflation  where
the vacuum energy driving inflation is   related to the same  $F$-term responsible   for the spectrum of supersymmetric particles  in gauge mediated supersymmetry breaking models. The inflaton is identified with  field responsible for the generation of the $\mu$-term.  
This opens the possibility of getting some knowledge about the low-energy supersymmetric theory from measurements of the cosmic microwave background radiation. Gravitinos do not  pose a cosmological problem, while the moduli
problem is ameliorated. 
\end{abstract}
\vskip2pc]

The existence of an inflationary stage during the evolution of the early
Universe is usually invoked to solve the flatness and the horizon problems of
the standard big  bang cosmology \cite{guth81}. During inflation the energy
density is dominated by the vacuum energy and comoving scales suffer an rapid
growth. As a result, any harmful topological defects left as remnants after
some Grand Unified phase transition, such as monopoles, are diluted.
Quantum fluctuations of the inflaton field imprint a
nearly scale invariant spectrum  of fluctuations on the background space time
metric. These fluctuations may be responsible for the generation of structure
formation. However, the level of density and temperature fluctuations observed
in the present Universe, $\delta\rho/\rho\sim 10^{-5}$, require the inflaton
potential to be extremely flat.
 For instance, in the chaotic inflationary scenario \cite{chaotic} where the
inflaton potential is $V=\lambda\phi^4$ and the condensate sits initially at
scales of order of the Planck scale, the dimensionless self-coupling $\lambda$
must be of order of $10^{-13}$ to be consistent with observations.
 The inflaton field must be coupled to other fields in order to ensure the
conversion of the vacuum energy into radiation at the end of inflation, but
these couplings must be very small  otherwise loop corrections to the inflaton
potential spoil its flatness. Such flatness is difficult
to understand in generic field theories, but
is quite common in supersymmetric theories\cite{ellis82,grisaru79}. 

Supersymmetry, then, might be expected to play a fundamental role during inflation \cite{lyth96},
but what this role might be will almost certainly depend on the
way in which supersymmetry is broken.  It is usually assumed that
supersymmetry is broken in a hidden sector and then transmitted
to observable fields either through gravitational\cite{su}\ or
gauge interactions\cite{gmsb}.  The cosmology of this latter class
of models has only been partially explored.
In this paper we will look for inflaton candidates in
this latter class of models,
those with gauge-mediated supersymmetry breaking
(GMSB) \cite{fc1}.

In this paper we would like to follow the minimal principle  that the inflaton field
should not be an extra degree of freedom inserted {\it ad hoc} in some (supersymmetric) theory of particle physics just to drive inflation. We will argue that this is indeed possible in the framework of GMSB models and show that a successful inflation model may be constructed. The energy density driving inflation may correspond to the $F$-term responsible for the spectrum of the superpartners in the low-energy effective theory. Moreover,
we will be able to identify the inflaton field with the same scalar responsible for the generation of the $\mu$-term present in the effective superpotential of the minimal supersymmetric standard model (MSSM), $\delta W\sim \mu H_U H_D$. 
This raises the possibility of someday connecting a theory which could be tested at accelerators
with measurements of the  temperature anisotropy in the
cosmic microwave background and related measurements of the two-point correlation function!
We will see that in such a picture, one can predict the fundamental scale
of supersymmetry breaking (which will turn out to be about $10^8$ GeV).  As it stands, our model
is not completely satisfactory.  As in virtually all models of
inflation based on supersymmetry, a fine tuning is required to
obtain a sufficiently small curvature for the potential near the
origin.  This tuning, at the level of a part in $10^2$,
must ultimately be given some deeper explanation.

In general, the effective low-energy theory that emerges from  GMSB
models contains soft SUSY-breaking mass terms for the scalar superpartners
which carry information about the scale and nature of the hidden-sector
theory. Typical soft breaking terms for sfermions
resulting from  GMSB models where SUSY is broken at the scale
$\Lambda$.  The magnitude of these terms
is given by  $\widetilde{m}\sim {{\alpha}\over{4\pi}}\Lambda$. This has
the interesting consequence that flavour-changing-neutral-current (FCNC)
processes are naturally suppressed in agreement with
experimental bounds. In the minimal GMSB models, a singlet field $X$ acquires a vacuum expactation value (VEV) for both its scalar and auxiliary components,
$\langle X\rangle$ and $\langle F_X\rangle$ respectively and 
the spectrum for the messangers is rendered non-supersymmetric.
Integrating out the messenger sector gives rise to gaugino
masses at one loop and scalar masses at two loops, with $\Lambda= F_X/X$.  
As we shall see, the energy density $V_0$ driving inflation may be identified with  the $F$-term responsible for the spectrum of supersymmetric particles, $V_0\sim F_X^2$. This is one of the key results of our paper.
  
Before launching ourselves into the inflationary aspects of this paper, let us discuss the issue of the generation of the $\mu$-term in the context of GMSB models. Leurer {\it et al.}, although in a different context, have suggested a $\mu$-term generation which is relevant for us \cite{le}. In addition to the usual MSSM fields, there is another singlet, $S$. As a consequence of discrete symmetries, the coupling $S H_U H_D$ is forbidden in the superpotential. There are, however, various higher dimension couplings which can drive $\langle S\rangle$. In particular, consider terms in the effective Lagrangian of the form
\begin{eqnarray}
&&\int d^2\theta\left(
\frac{1}{M_p^p} X S^{2+p}+\frac{1}{M_p^m} S^{m+3}+\frac{1}{M_p^n}S^{n+1}H_U H_D\right)\nonumber\\
&+& \frac{1}{M_p^2}\int d^4\theta\:X^{\dagger}X S^{\dagger}S,
\end{eqnarray}
where $M_p\simeq 1.2\times 10^{19}$ GeV is the  Planck mass.  This structure can be enforced
by discrete symmetries.
The first and the fourth terms can contribute to the effective negative curvature terms to the $S$ potential. For example, if $p=m=2$, and  $n=1$, then the $\mu$-term turns out to be of the order of $\sqrt{F_X}$ (times powers of coupling constants).
%Besides generating a $\mu$-term, this mechanism can also give rise to a nearly
%
%vanishing  $B$-term, {\it i.e.}
% the soft supersymmetry breaking
%term $m_3^2 H_uH_d$ in the Higgs potential.
% Boundary conditions equal to zero for
% bilinear (and trilinear) soft parameters at
%the messanger scale makes
%the GMSB models free from the supersymmetric CP problem and highly predictive \cite{bor}. 

The field $S$, although very weakly coupled to ordinary matter, may play a significant role in cosmology. We will devote the rest of the paper to explore the cosmological implications of such a field and to show that a succesfull inflationary scenario may be constructed out of the potential for the field $S$.

Let us suppose that    the field $X$ has acquired a vacuum expactation value (VEV) for both its scalar and auxiliary components and restrict ourselves to the case  $p=m=2$, and  $n=1$. The potential along the real component of the  field $S$ reads
\begin{eqnarray}
V(S) &\sim & \beta^2 F_X^2-\alpha\frac{F_X^2}{M_p^2}S^2-\beta\frac{F_X}{M_p^2}S^4\nonumber\\
&+& \beta^2 \frac{X^2}{M_p^4}S^6 - \beta\frac{X}{M_p^4} S^7+\frac{1}{M_p^4}S^8,
\end{eqnarray} 
where we have explicitly written the coupling constants. The true vacuum is at $\langle S\rangle^4\sim \beta F_X M_p^2$, such that $\mu\sim \langle S\rangle^2/M_p\sim \sqrt{\beta F_X}$. Notice that 
we have added the constant $\sim F_X^2$ in such  a way that the the cosmological constant in the true vacuum is zero,   $V(\langle S\rangle)=0$.
As we will see, it is necessary that $\alpha$ be small, of order $10^{-2}$.  On the other hand, in the framework
of supergravity, it is well-known that there are contributions to the mass of particles during
inflation of order $H$ from the terms $S^{\dagger}S$ which must be present in the Kahler potential.
These can be cancelled by terms of the form $S^{\dagger}S X^{\dagger}X$.  This is the fine tuning we have alluded
to earlier.  It is a generic feature of models based on supersymmetry\cite{mass}. One
positive feature of these models is that the curvature at the minimum
of the potential is automatically far larger than the curvature near
the origin.  This is important for reheating.

Around $S=0$ we may simplify the potential as
\begin{equation}
V(S)\simeq V_0-\frac{m^2}{2}S^2-\frac{\lambda}{4}S^4,
\end{equation}
where $V_0\sim \beta^2F_X^2$,  $m^2\sim\alpha\frac{F_X^2}{M_p^2}$ and 
$\lambda\sim 4\beta \frac{F_X}{M_p^2}$.
If the $S$-field starts sufficiently close to the origin the system may inflate. $S$ rolls very slowly toward $S=\langle S\rangle$ during inflation. Its potential energy dominates the energy density of the Universe driving a nearly constant expansion rate
\begin{equation}
H^2(S)=\frac{8 \pi \:V(S)}{3\:M_p^2}\simeq\frac{8 \pi \:V_0}{3\:M_p^2}.
\end{equation}
In the slow-roll approximation, we may neglect $\ddot{S}$ in the equation of motion so that $\dot{S}\simeq-V^\prime/3H$. This approximation is consistent as long as
$V^\prime,V^{\prime\prime}\ll V$ for $S\sim 0$. 
Notice that the quadratic and the quartic terms in the expression (4) become comparable for $S_*\sim\sqrt{F_X}$. Since this value   is much smaller than $\langle S\rangle$ and all the dynamics giving rise to density perturbations
takes place  in the vicinity of the origin, we prefer not to   neglect the quartic term in the rest of our analysis. 

During the slow-roll phase, when $S$ is near to the origin, the cosmic scale factor may grow by $N(S)$ e-foldings:  
\begin{eqnarray}
\label{n}
N&\simeq&\frac{8\pi}{M_p^2}\int_{S_N}^{S_e}\:\frac{V_0}{-V^\prime(S)}
\simeq
\frac{2\pi V_0}{m^2 M_p^2}\nonumber\\
&\times&\left[1-4\:{\rm ln}\left(\frac{S_N}{S_*}\right)\right] ,
\end{eqnarray}
where $S_e$ denotes the value of the field when inflation ends. Successfull inflation requires $N\simeq 60$. We can express now  $S_N$ in terms of $S_*$:
$S_N\simeq S_*{\rm exp}\left[1/4-N(m^2M_p^2/8\pi V_0)\right]$.  
This expression is valid as long as $S_N<S_*$. 
The value of the field $S_e$ at which inflation ends corresponds to the moment when the parameter $\epsilon=(M_p^2/4\pi)\left(H^\prime(S)/H(S)\right)^2$ becomes smaller than 1. One can define  another fundamental parameter, $\eta=(M_p^2/4\pi)\left(H^{\prime\prime}(S)/H(S)\right)$. 
In the slow-roll approximation, these parameters reduce to 
 $\epsilon=(M_p^2/16\pi)(V^\prime/V)^2$ and $\eta=(M_p^2/8\pi)
(V^{\prime\prime}/V)$ and the slow-roll approximation is valid as long as $\epsilon,\eta\ll 1$. In the model under consideration, however, the slow-roll limit  breaks down well before the end of inflation. Using the slow-roll solution for $\epsilon$ results in understimating the value $S_e$ at which inflation ends \cite{kinney}, the correct value being $S_e$ of the order of $\langle S\rangle$. Even though the correct value of $\eta$ becomes large before the end of inflation, the slow-roll approximation is valid for $S\ll \langle S\rangle$, which is the region where observable parameters must be computed \cite{kinney}. 
In this region one may compute the power spectrum, which is the Fourier transform of the two-point density autocorrelation function. It has the primordial form $P(k)\propto k^n$. where $k$ is the amplitude of the Fourier wavevector and $n$ denotes the spectral index.
Fluctuations arise due to quantum fluctuations in the scalar field $S$. The measurement of the quadrupole anisotropy in the cosmic
microwave background radiation detected by COBE 
 allows us to fix the parameters of the model:
\begin{equation}
\left(\frac{\Delta T}{T}\right)=\sqrt{\frac{32\pi}{45}}\frac{V_0^{3/2}}
{V^\prime(S_N) M_p^3}\simeq
 \frac{\beta^{7/2}}{\alpha^{3/2}}\frac{\sqrt{F_X}}{M_p}{\rm e}^{-\frac{1}{4}+N|\eta |}.
\end{equation}
Imposing $\left(\frac{\Delta T}{T}\right)\simeq 6\times 10^{-6}$ and fixing $\beta\sim 1$, $\alpha\sim 10^{-2}$
 and $\sqrt{F_X}\sim 10^8$ GeV gives 
$|\eta |\sim 10^{-1}$. It is easy to verify that for such a value of $|\eta|$ the condition $S_N<S_*$ we have used in eq. (7) is indeed satisfied. 
At a first sight the value $\sqrt{F_X}\sim 10^8$ GeV might seem too large in the framework of GMSB models. However, the spectrum of the superparticles depends on  the  ratio $\Lambda= F_X/X$. The latter is fixed to be relatively small and in the range $(10-10^3)$ TeV. This may be easily accomplished if  $X$
acquires a large vacuum expectation value via nonrenormalizable operators. 
Notice that $\sqrt{F_X}\sim 10^8$ GeV corresponds to a very low value of the Hubble parameter during inflation, $H\sim$ 1 MeV.
Note also that it implies that there is a small dimensionless number in the coupling of $S$ to the Higgs.  This number
is comparable to the electron Yukawa coupling. 

The COBE satellite  measured fluctuations in the cosmic microwave background with a spectral index $n = 1.2 \pm 0.6$ (at 2$\sigma$ level). In our model
the  spectral index
$n\simeq 1-2|\eta |\sim 0.8$ is noticeably smaller than  1 and gives rise to a red spectrum.
 The amplitude of the gravitational waves produced by quantum fluctuations is far too small to be detected since the variation of the field during inflation is much smaller than $M_p$ \cite{grav}.

Let us briefly address the issue of the initial condition for the field $S$.  We have assumed that underlying the
model are discrete symmetries under which $S$ transforms non-trivially.  As a result,
$S=0$ is a special point, and it is natural that $S$ may ``sit'' at this point initially.
This despite the fact that it is very weakly coupled to ordinary matter, and might not
be in thermal equilibrium.
Many models of slow-rollover inflation require a fine-tuning in the initial value for the field to be successfull and the smaller is the scale of inflation the more severe is the fine-tuning \cite{tu}. From Eq. (7) we may infer that,  in order to achieve the 60 or so e-foldings of inflation required, the initial value of the scalar field must be less than about $2\times 10^{5}\sqrt{\alpha/\beta} \left(\sqrt{F_X}/10^{8}\:{\rm GeV}\right)$ GeV.
As a result, only regions  where the initial value of the field is small enough will undergo inflation. These regions have grown exponentially in size and they should occupy most of the physical volume of the Universe. We notice that the small value of the field  is not spoiled by quantum fluctuations which are of the order of $H/2\pi\sim 4\times 10^{-4}\beta \left(\sqrt{F_X}/10^{8}\:{\rm GeV}\right)^2$ GeV. Thermal fluctuations might spoil such a localization since $
\langle S^2\rangle_T^{1/2}\sim T\sim \sqrt{F_X}$. However, the inflaton field is so weakly coupled (its
couplings are all suppressed by powers of $M_p$ that it is not in thermal contact with the rest of the Universe \cite{tu,ross}.

After inflation ends, the $S$-field starts oscillating around the minimum of its potential and the vacuum energy that drives inflation is converted into coherent scalar field oscillations. The Universe undergoes a period of matter domination. 
Reheating takes place when $S$ decays into light fields, which will eventually thermalize and give rise to a thermal bath of radiation. The reheating temperature $T_{R}$ is determined by the decay width of the 
scalar oscillations $\Gamma_S$, $T_R\sim 0.1\sqrt{\Gamma_S M_p}$ \cite{rocky}. At the minimum of its potential, the scalar field has a mass
squared $m_S^2=V^{\prime\prime}(S)\simeq 10 \frac{F_X^{3/2}}{M_p}$. For example, for $\sqrt{F_X}=10^8$ GeV, $m_S\sim 1$ TeV. The scalar oscillations may decay into light Higgsinos $S\rightarrow \psi_{H_u}\psi_{H_d}$ with a rate
$\Gamma_S=g^2 m_S/4\pi$ with $g\sim\frac{\mu}{\sqrt{F_X}}\frac{\langle S\rangle}{M_p}$ and we have taken into account that $\mu\ll \sqrt{F_X}$. The  resulting reheating temperature is then $T_R\sim 10^{-2} \mu F_X^{1/8}M_p^{-1/4}\simeq 10^2\left(\mu/
10^3\:{\rm GeV}\right)\left(\sqrt{F_X}/
10^8\:{\rm GeV}\right)^{1/4}$ MeV, which is large enough to preserve the classical cosmology beginning with the era of nucleosynthesis. It seems difficult, however, to push the reheating temperature above the electroweak scale. Thus, it appears that electroweak baryogenesis is not a viable option
for the generation of the baryon asymmetry. On the other hand,
the decays of the inflaton themselves might be responsible for the baryon asymmetry \cite{b}.  
The couplings by which the inflaton decays may contain CP-violation and baryon number violation. In order to produce a baryon asymmetry, we must have baryon number violating operators in the Lagrangian, such as $\delta W\sim \left(\frac{S}{M_p}\right)\bar{U}\bar{D}\bar{D}$. 
The presence of such operator is compatible with the stability of the proton and the experimental absence of neutron-antineutron oscillations. We can estimate the baryon asymmetry produced by the inflaton decay simply. We assume that the amount of baryon number produced per decay is $\varepsilon$. $\varepsilon$ is the product of CP-violating phases $\delta$ times the ratio of the baryon number violating decay rate over the total decay rate $\Gamma_{ B}/\Gamma_{tot}\sim 10^{-2} \left(\mu/\sqrt{F_X}\right)^2\left(m_S/\langle S\rangle\right)^2$. 
 The number of massless particles produced per decay is $\sim m_S/T_R$. Plugging in the expected values  of the inflaton mass and the reheating temperature for $\sqrt{F_X}\sim 10^8$ GeV, we find a baryon to entropy ratio $B\sim  10^{-8}\:\delta$, which is compatible with the observed value for $\delta\sim 10^{-2}$. 

Another remarkable effect of  a late period inflation is that the gravitino problem is solved. As is well known, in GMSB scenarios the gravitino is the lightest supersymmetric particles with mass $m_{3/2}\sim F_X/M_p$ (of order of  1 MeV in our case).  If a stable gravitino is thermalized in the early Universe and not diluted by any mechanism, its mass density may exceed the closure limit $\Omega_{3/2}\la1$. Since the number density of gravitinos is fixed once they  are thermalized, the above argument sets a stringent  upper bound on the gravitino mass,  $m_{3/2}\la 2$ keV (without dilution) \cite{pargel}. 
However, gravitinos are efficiently diluted during the inflationary stage driven by the field $S$ and they are not produced in the subsequent stage of reheating. Indeed,  light gravitinos (or, better to say, the longitudinal components of them) may be regenerated during reheating  either by the  decays of sparticles (or particles in the messanger sector)    or by scatterings processes. However,  the first mechanism requires the reheat temperature to be at least of order of  the typical sparticle mass,  $\widetilde{m}\sim$ 100 GeV, and scattering processes     
regenerate the light gravitinos only   if $T_R\ga 10^2\left(m_{3/2}/1\:{\rm MeV}\right)$ TeV \cite{murayama}. Since in our scenario the reheat temperature turns out to be very low, $T_R\sim 10^2$ MeV, we may safely conclude that gravitinos were not populating the  Universe at the beginning of the radiation era:   the  gravitino problem is solved by the late stage of inflation. 

In models of gauge mediation, if we assume that the underlying theory is a string theory,
the cosmological moduli problem is even more severe than in the usual supergravity scenarios \cite{moduli}. 
String moduli  acquire  a mass comparable to $m_{3/2}$ and are stable on cosmological time scales.
As a result, their energy density is a severe problem \cite{murayama}. However,  this problem is
somewhat ameliorated in our scenario. The  period of inflation  driven by
the field $S$ may take place
at a sufficiently   late stage of the Universe,   $H\la m_{3/2}$, that
the number density of string moduli is reduced by a factor ${\rm exp}(-3N)$ and  by the subsequent entropy production at the reheat stage \cite{randall}. It is an attractive
feature of the present scenario that this is possible.
However, this is not enough, since the minimum
of the moduli potential, generically, will be shifted by an amount of order $M_p$ during
inflation, as a result
of couplings of the moduli
\begin{equation}
\int d^4 \theta X^{\dagger} X f({\cal M}) + S^{\dagger} S
g({\cal M}).
\end{equation}
So it is probably necessary to find symmetry reasons
that the minima coincide to a high degree of accuracy\cite{mass}.

There has been much discussion of the $\mu$ problem in the context of gauge
mediation, and it is not clear what is the most satisfactory
solution.  It is appealing that the solution in \cite{fc1}\ also provides
an interesting candidate inflaton.  We have seen that provided $F_X$
is of suitable size, and provided that certain parameters of order
one take suitable values, inflation can occur.

The work of M.D. was supported in part by the U.S. Department of Energy.     A.R.
is supported by the DOE and NASA
under grant NAG5--2788.  MD thanks the Institute for Theoretical
Physics at Stanford University for its hospitality, and Scott Thomas for discussions.

\def\NPB#1#2#3{Nucl. Phys. {\bf B#1} (19#2) #3}
\def\PLB#1#2#3{Phys. Lett. {\bf B#1} (19#2) #3}
\def\PLBold#1#2#3{Phys. Lett. {\bf#1B} (19#2) #3}
\def\PRD#1#2#3{Phys. Rev. {\bf D#1} (19#2) #3}
\def\PRL#1#2#3{Phys. Rev. Lett. {\bf#1} (19#2) #3}
\def\PRT#1#2#3{Phys. Rep. {\bf#1} (19#2) #3}
\def\ARAA#1#2#3{Ann. Rev. Astron. Astrophys. {\bf#1} (19#2) #3}
\def\ARNP#1#2#3{Ann. Rev. Nucl. Part. Sci. {\bf#1} (19#2) #3}
\def\MPL#1#2#3{Mod. Phys. Lett. {\bf #1} (19#2) #3}
\def\ZPC#1#2#3{Zeit. f\"ur Physik {\bf C#1} (19#2) #3}
\def\APJ#1#2#3{Ap. J. {\bf #1} (19#2) #3}
\def\AP#1#2#3{{Ann. Phys. } {\bf #1} (19#2) #3}
\def\RMP#1#2#3{{Rev. Mod. Phys. } {\bf #1} (19#2) #3}
\def\CMP#1#2#3{{Comm. Math. Phys. } {\bf #1} (19#2) #3}

\vskip 3 cm

\end{document}